\documentclass[11pt,revtex4]{article}
\usepackage{graphics}

\begin {document}
\title{\bf \  N-dimensional gravitational
collapse of Type I matter fields with non-zero radial pressure}

\author {Sanjay B.~Sarwe \footnote{email:sbsarwe\_ngp@sancharnet.in}\\
  Department of Mathematics, S. F. S. College,
  \\ Seminary Hill,
   Nagpur-440 006, India  \\
 R. V. Saraykar  \footnote{email:r\_saraykar@rediffmail.com} \\
 Department of Mathematics, Nagpur University \\
 Campus, Nagpur-440 033, India}
\date{}
\maketitle
\abstract
 { We study the gravitational collapse of Type
I matter field in $N$ dimensional spacetime with radial pressure
$p_{r}$ as a function of $r $. We find that for a given smooth
initial data set satisfying physical requirements, naked
singularities exist for spacetime dimensions $N=4$ and $5$ while
for $N \geq 6$ Cosmic Censorship Conjecture withholds its ground.
We, also, study the collapse with linear equation of state and
find that, similar to dust collapse with appropriate choice of
initial data naked singularities occur in all dimensions.} \\


\section* {I. Introduction}
In the recent paper \cite{rgpj-1} R. Goswami and P. S. Joshi
pointed out that the occurrence of naked singularity (NS) as end
state of dust collapse beginning from smooth initial data with
physically relevant restrictions, can be removed when one goes to
higher dimensional space-time. This has restored Cosmic Censorship
Conjecture (CCC)(articulated by Roger Penrose \cite{rp}) in dust
collapse for sufficiently high dimension of the space-time. Many
other cases of the dust collapse have been studied by various
authors \cite{bd,kd}.

A lot of importance is being given to the physically relevant
models such as Type I matter fields that include most of the known
physical forms of matter like dust, perfect fluid, etc. The
authors in  \cite {jd, jo, jr} have discussed the role of initial
data in spherically symmetric gravitational collapse for Type I
matter fields. Recently, in the paper \cite{amrgpj} A. Mahajan et
. al have shown the restoration of CCC in Einstein cluster
spacetime for spacetime dimensions $N \geq 6$ with non-zero
tangential pressure $p_{\theta}$ and radial pressure $p_{r} = 0$.

 F.I. Cooperstock and et al. \cite {cjjs} have shown that
if the radial pressure $p_{r}$ is positive , $r>0$ then
non-central singularity is covered by the horizon irrespective of
the sign of tangential pressure $p_{\theta}$.

In recent years, special attention is given to higher dimensional
gravitational collapse since the development in string theory
indicate that gravity may be a higher dimensional interaction.
Hence, it would be interesting to know the status of CCC in higher
dimension spacetime. These studies may help in possible
appropriate mathematical formulation of the conjecture or assist
in providing any possible proof of CCC.

We take up the problem of Type I matter field collapse with radial
pressure $p_{r}$ as a function of $r$ and arbitrary tangential
pressure $p_{\theta}$. The apparent horizon for the evolving
collapse has been studied and it is found that the final state of
collapsing cloud results in the black hole (BH) for the dimension
of the spacetime $N \geq 6$.

Introduction of equation of state has importance as it makes the
collapsing models more physically relevant. Does the linear
equation of state $p = k \rho$ stimulates the outcome of NS/BH
phases as end state of gravitational collapse? We study, this case
in section V.

In section II, we briefly summarize the field equations and state
the conditions on the initial data under which the collapse
evolves. The evolving field equations with $p_{r} = p_{r}(r)$ have
been illustrated in section III. In section IV, the behaviour of
the apparent horizon is studied. The case with linear equation of
state has been analyzed in section V. Finally, we make some
concluding remarks.

\section* {II. Field equations and initial data }
 The general spherically symmetric metric describing $N$ dimensional
 space-time geometry within the collapsing cloud can be described in
 comoving coordinates $(t, r, \theta, \phi)$ by
\begin{equation}
 ds^2 = - e^{2 \nu(t,r)}dt^2 +  e^{2 \psi(t,r)}dr^2 + R^2(t,r) d
\Omega^2_{N-2} \label {m01}
\end{equation}
where
\begin{eqnarray*}
 d\Omega^2_{N-2}=\sum_{i=1}^{N-2} \left[\prod_{j=1}^{i-1}
 \sin^2(\theta^{j})\right]\;(d\theta^{i})^2 \nonumber
\end{eqnarray*}
 is the
metric on an $(N-2)$ sphere. The stress-energy tensor for Type I
field in a diagonal form is  given by \cite {he}
\begin{equation}
 T{^{t}_{t} } = -\rho, \; T{^{r}_{r} } = \; p_{r},  \;
 T{^{\theta}_{\theta}}=T{^{\phi}_{\phi} } = \; p_{\theta} \label {em02}
\end{equation}
The quantities $\rho$, $p_{r}$ and $p_{\theta}$ are the energy
density, radial and tangential pressures respectively. We take the
matter field to satisfy weak energy condition i.e. the energy
density measured by any local observer be non-negative, so for any
vector $V^{i}$, we must have, $T_{ik}V^{i} V^{k} \geq 0$ which
means ${{\rho}{\geq}}0$ ;  ${\rho+p_{r} {\geq}} 0$ and
${\rho+p_{\theta} {\geq}} 0$ .

Einstein field equations for the metric (1) are described by
\begin{equation}
\rho = \frac{(N-2) F'}{2 R^{N-2}R'} \label {fe03}
\end{equation}
\begin{equation}
 p_{r} = - \frac{(N-2) \dot{F}}{2 R^{N-2} \dot{R}} \label {fe04}
 \end{equation}
\begin{equation}
\nu' = \frac{(N-2) (p_{\theta}-p_{r}) \; R'}{(\rho + p_{r}) R } -
\frac{p'_{r}}{\rho + p_{r}} \label {fe05}
\end{equation}
\begin{equation}
 - 2 \dot{R'} + R'\frac{\dot{G}}{G} + \dot{R}
\frac{H'}{H} = 0 \label {fe06}
\end{equation}
\begin{equation}
\hspace{.5in} G - H = 1 - \frac{F}{R^{N-3}} \label {fe07}
\end{equation}
where we have defined $ G(t,r) = e^{-2 \psi } (R')^2 $ and $
H(t,r) = e^{ -2 \nu }( \dot{R})^2$.

The arbitrary function $F(t,r)$ represents the total mass in a
shell of collapsing cloud of comoving radius $r$. The weak energy
conditions imply $ F \geq 0 $. The regularity at the initial epoch
$t=t_{i}$ is preserved by $F(t_{i},0) = 0$ i.e. the mass function
should vanish at the centre of the cloud. The space-time density
singularity is due to $R = 0$ and $R' = 0$, the later one is due
to shell-crossings and can possibly be removed from the space-time
\cite {r1}, hence we consider here only a physically relevant
singularity $R=0$ known as shell-focusing singularity where all
matter shells collapse to a zero physical radius. The scaling
freedom available for the radial co-ordinate $r$ is being used to
introduce the function $v(t,r)$ by the relation
\begin{equation}
R(t,r) = r v(t,r), \label {sf08}
\end{equation}
this relation is due to the definition $ v(t,r) = R / r $ \cite
{rg2}. We have $v(t_{i},r) = 1$; $ v(t_{s}(r), r) = 0 $ and for
collapse $ \dot{v} < 0 $.  The time $ t = t_{s}(r) $ corresponds
to the shell-focusing singularity $ R = 0$. The six arbitrary
functions of the shell radius $r$ as given by $ \nu(t_{i},r) =
\nu_{o}(r)$, \ $\psi(t_{i},r) = \psi_{o}(r)$, \ $R(t_{i},r) = r$,
\ $\rho(t_{i},r) = \rho_{o}(r)$, \ $p_{r}(t_{i},r) =
p_{r_{o}}(r)$,
 \ $p_{\theta}(t_{i},r) = p_{\theta_{o}}(r)$ evolve the dynamics of
the initial data prescribed at the initial epoch $ t = t_{i}$. We
have a total of five equations with seven unknowns, namely $ \rho,
p_{r}, p_{\theta}, \nu, \psi, R$ and $F$ giving us the freedom of
choice of two functions. Selection of these two free functions,
subject to the given initial data and the weak energy condition
above, determines the matter distribution and the metric of the
spacetime and thus, leads to a particular time evolution of the
initial data. The existence and uniqueness of solution of the
system of field equations with above mentioned initial data  has
been discussed by Joshi and Dwivedi \cite {dj}. The solution
continues to exist in the neighbourhood of the singularity given
by $R=0$.

\section* {III. Radial pressure $p_{r}=p_{r}(r)$ }
We consider radial pressure $P_{r}$ as a function of $r$.
 The two allowed free functions $p_{r}$ and $\nu(t,r)$ are chosen as follows:
\begin{equation}
p_{r} = p_{r}(r), \hspace{.1in} \nu(t,r) = c(t) + {\eta} (R)
\label {pr09}
\end{equation}
Now, let us see, how the Einstein's field equations react to this
choice which will decide the end state of collapse. As $p_{r}$ is
a function of $r$ alone, integrating equation (\ref{fe04}), we
obtain
\begin{eqnarray*}
F(t,r) = - \frac{2 \; p_{r} R^{N-1}} {(N-1) (N-2)}+ z(r)
\end{eqnarray*}
where $z(r)$ is another arbitrary function of $r$. We need to
choose $z(r)$, \\
$ z(r) = [{4}/{(N-1) (N-2)}] r^{N-1} p_{r}(r) $
so that the mass function takes the form
\begin{equation}
F = \frac{2 \; p_{r} ( 2 r^{N-1} - R^{N-1} )}{(N-1) (N-2) } \label
{mas10}
\end{equation}
that satisfies the regularity conditions at the initial epoch $
F(t_{i}, r) =  2 \; p_{r} r^{N-1}/(N-1) (N-2) $ and $ F(t_{i}, 0)
= 0 $. From equation (\ref{mas10}), it is evident that
non-negativity of radial pressure maintains non-negativity of mass
function throughout the gravitational collapse ie. $ F \geq 0 $
provided $ p_{r} \geq 0 $. Next, equation (\ref{fe03}) becomes
\begin{equation}
\rho =  \frac{(N-1) \; p_{r} \; ( 2 r^{N-2} - R^{N-2} R')
  + p'_{r}  \ (2 r^{N-1} - R^{N-1})}{ (N-1) \; R^{N-2} R'} \label {den11}
\end{equation}
At the initial epoch $ t = t_{i}$,
\begin{equation}
 \rho_{o} = p_{r} + \frac{1}{N-1} r p_{r}'  \geq 0   \label {dei12}
\end{equation}
 , an all important relation between energy
density, radial pressure and pressure gradient.
 In the limit of approach
to the singularity ie. as $ r \rightarrow 0 $ and $ t \rightarrow
t_{s}$
\begin{equation}
 \rho = \frac{ p_{r}(0) (2 - v^{N-1})} {v^{N-1}} ,  \label {des13}
\end{equation}
the density blows up at the singularity. Equation ( \ref{fe07} )
can be written in the form
\begin{equation}
 G = e^{2 \ \eta(R)} d(r)  \label {g13}
\end{equation}
with $d(r) = 1 + r^{2} b(r)$ where $b(r)$ is the energy
distribution function for the collapsing shells. Using equation
(\ref{mas10}) in equation (\ref{fe05}), we obtain
\begin{equation}
( \rho + p_{r} ) \ \eta_{,R} \ R' R  = (N - 2) ( p_{\theta} -
p_{r} ) R' - R \ p'_{r}  \label {dpr14}
\end{equation}
The tangential pressure blows up in the limit of approach to the
singularity as the density diverges at the singularity and  we
observe that all of initial data is not independent at the initial
epoch $t = t_{i}$
\begin{equation}
( \rho + p_{r} ) \ \eta_{,R} \ r  = (N-2) ( p_{\theta} - p_{r} ) -
r \ p'_{r} \  \label {dpr15}.
\end{equation}
 Using the initial data prescribed at the initial epoch $ t = t_{i}$
 to evolve the collapse, we
 can integrate equation (\ref{dpr15}) and obtain,
 \begin{equation}
\eta(R) =  \int_{0}^{R} \frac{(N-2) (p_{\theta_{o}} - p_{r_{o}}) -
R p_{r_{o}}' } {R ( \rho_{o} + p_{r_{o}})}dR   \label {eta16}
\end{equation}

so, velocity distribution function $ \eta(R)$ is determined by
smooth functions $ p_{r}$ and $ p_{\theta}$. Using equations
(\ref{mas10}) and (\ref{g13}), equation (\ref{fe06}) can be put in
the form
\begin{eqnarray}
 R^{(N-2)/2} {\dot{R}} = - \ a(t) \ e^{\eta} \
   { \sqrt{[ 1 + r^2 b(r)] R^{N-3} e^{2 \eta}- R^{N-3}  +
\frac{2 p_{r} \; ( 2 r^{N-1} - R^{N-1})}{(N-1) (N-2)}  }}   \label
{r17}
\end{eqnarray}
where $a(t)$ is a function of time and we choose $a(t) = 1$ by
suitable scaling of time co-ordinate.\\
We define
\begin{equation}
h(R) = \frac{ e^{2 \eta(R)} - 1} {R^2} = 2 \ g(R) + O(R^2) \label
{h18}
\end{equation}
Using this definition, equation (\ref{r17}) becomes
\begin{equation}
    v^{(N-3)/2} \ {\dot{v}} = - e^{\eta}
   \sqrt{v^{N-3} b(r) e^{2 \eta} +  v^{N-1} h(rv) +
\frac{2 p_{r}\; ( 2 - v^{N-1}) }{(N-1) (N-2)}  }  \label {v19}
\end{equation}
Integrating above equation, we obtain
\begin{equation}
 t(v,r) = \int_{v}^{1} \frac{v^{(N-3)/2} dv} { e^{\eta}
 \sqrt{ v^{N-3}  e^{2 \eta} b(r) +   v^{N-1} h(rv) + \frac{2 \; p_{r}\;
  ( 2 - v^{N-1})} {(N-1) (N-2)} } }   \label {v20}
\end{equation}
where the variable $r$ is treated as a constant. Further, the time
taken for the central shell to reach the singularity is given by
\begin{equation}
 t_{s_{o}} = \int_{0}^{1} \frac{v^{(N-3)/2} dv} {
 \sqrt{ v^{N-3}  b(0) +   v^{N-1} h(0) + \frac{2 \; p_{r}(0) \;
  ( 2 - v^{N-1})} {(N-1) (N-2)} } }  \label{ts21}
\end{equation}
and $t_{s_{o}}$ is well defined provided $ [ v^{N-3}  b(0) +
v^{N-1} h(0) + [2 \; p_{r}(0) \;
  ( 2 - v^{N-1})] / [(N-1) (N-2)] ] > 0 $. The Taylor expansion
  of $ t(v,r) $ around the center $r=0$ is given by
\begin{equation}
t(v,r) = t(v,0) + r \; \chi_{p}(v) + O(r^2)    \label {tvr21a}
\end{equation}
The time taken for other shells close to center $r=0$ to reach the
singularity as $ t\rightarrow t_{s}$ can be determined from
\begin{equation}
t_{s}(r) = t_{s_{o}} + r \; \chi_{p}(0) + O(r^2) \label {tsr22}
\end{equation}
 and we have
\begin{eqnarray}
 \chi_{p}(v) = - \ {\frac{N-3}{2}} \
       \int_{v}^{1} {  \frac{v^{(N-3)/2}
\left[ v^{N-3}  b'(0) +   v^{N-1} h_{,r}(0) + \frac{2 \; p_{r}'(0)
\;
  ( 2 - v^{N-1})} {(N-1) (N-2)} \right]}{\left[v^{N-3}  b(0) +
    v^{N-1} h(0) + \frac{2 \; p_{r}(0) \;
  ( 2 - v^{N-1})} {(N-1) (N-2)}  \right]^{3/2}} }dv. \label {chi24}
\end{eqnarray}
It is clear that $\chi_{p}(v)$ depends on the functions
$p_{r}(0),h(0)$ and $b(0)$ and these functions, in turn, depend on
the given set of density, tangential pressure and velocity
function $\nu$. For any constant v surface, we have $dv = 0$ which
implies
 $\dot{v} \;  \chi_{p}(v)  = -  v'$ .

 Now, equation (\ref {v19}) takes the form
\begin{equation}
v^{(N-3)/2} v' = \left[ \chi_{p}(v)  \right] \; D(0,v) \label
{cv25}
\end{equation}
where
\begin{equation}
D(0,v)=
   \sqrt{v^{N-3}  b(0)+  v^{N-1} h(0) +
\frac{2 \; p_{r}(0) \; ( 2 - v^{N-1}) } {(N-1) (N-2)} } \label
{cv26}
\end{equation}
From equations (\ref{dpr15}) and  (\ref{dei12}), it is evident
that all of initial data is not independent. Hence, we conclude
that $b(r)$, radial and tangential pressures $p_{r}$ and
$p_{\theta}$ will be sufficient to form initial data set for the
collapse. Now, we assume that the initial data functions $b(r)$,
$p_{\theta}(r,v)$ and $p_{r}(r)$ are at least $C^2$ functions and
that satisfy the regularity conditions at the center $r = 0$ ie.
$p'_{{r}_{o}}(0)=p'_{{\theta}_{o}}(0) = 0 $ and $ p_{{r}_{o}} (0)
- p_{{\theta}_{o}}(0) =0$.

Using equation (\ref{cv25}), we can write
\begin{equation}
\chi_{p}(0) = \lim_{v \rightarrow 0} \chi_{p}(v) = \lim_{ v
\rightarrow 0} \frac{ v^{\frac{N-3} {2} } v'} { D(0,v) }
\label{ch26}
\end{equation}
 Since the function $D(0,v)> 0$ and $v \in
[0,1]$ and if the above limit exists, then the sign of
$\chi_{p}(0)$ as positive or negative shall depend on the
behaviour of $v'$ . Hence, $v' > 0$ may give rise to $\chi_{p}(0)
> 0$ as limit of sequence of positive terms can not be negative.
If $\chi_{p}(0) = 0$ then we need to consider the higher
derivative of $t(v,r)$ at $r=0$ for similar analysis.

\section* {IV. Study of apparent horizon}
The end state of gravitational collapse, in terms of either a
naked singularity or a black hole, is determined by the causal
structure of non-spacelike curves in the neighbourhood of
singularity. The singularity is visible if there exist future
directed families of non-spacelike curves emanating from the
singularity that reach the far away observers in the future, and
which have past end point at the singularity. Otherwise event
horizon forms early enough to cover the singularity and thus, the
end state of collapse is a black hole. To determine this
characteristic of singularity, we study the behaviour of apparent
horizon which is the boundary of trapped surfaces as the collapse
evolves. This boundary of the trapped region of the space-time
within the collapsing cloud is  given by the equation,
\begin{equation}
\frac{F} {R^{N-3}} =1   \label {ap01}
\end{equation}
and which is the equation for the apparent horizon, for details,
we refer the reader to \cite{amrgpj}. Firstly to know about the
NS/BH phases for any $r > 0$ along the singularity curve $t=
t_{s}(r)$. For example, from above equation along the singularity
curve (which corresponds to $R=0$), for any $r > 0$, the mass
function $F(t,r)$ takes the form
\begin{eqnarray*}
F(t,r) = c_{1} (2 - v^{N-1}),
\end{eqnarray*}
for a specific choice of $r$, whereas the physical area radius $R
(=c_{2} v) \rightarrow 0$ as $v \rightarrow 0$ where $c_{1}$ and
$c_{2}$ are constants. Therefore, trapped surfaces form that cover
the neighbourhood of the center before the development of
singularity.

Using equation (\ref {ap01}), the apparent horizon $v_{Nah}$ is
obtained through the equation
\begin{equation}
v^{N-1} + \frac{ (N-1) (N-2) } {2 r^2 p_{r} } v^{N-3}- 2 = 0.
\label {app2}
\end{equation}
 We know from
above that all the points $r>0$ on the singularity curve are
already covered and the behaviour of apparent horizon is being
studied close to the central singularity at $r=0,R=0$, therefore,
the values of $r$ in this analysis must be sufficiently small.
With this background, the equation (\ref{app2}) is a polynomial
equation in $v$. For dimensions $N = 4$ and $N=5$ the $v_{Nah}$
are determined as positive roots of above polynomial equation and
we obtain
\begin{eqnarray}
v_{4ah} = \frac{A_{1}^{1/3}} { r \; p_{r}^{1/3} } -
\frac{A_{1}^{-1/3}} { r \; p_{r}^{2/3} } ; \; \hspace{.5in}
 A_{1} = r^3 p_{r}
+ \sqrt{ \frac{ 1 + r^6 \; p_{r}^{3} } {p_{r}} }. \label {v4ah}
\end{eqnarray}
and
\begin{eqnarray}
v_{5ah} =  \frac{ \sqrt{ p_{r} \left[ \; -3 + \sqrt{  9 + 2 r^4 \;
p_{r}^2} \;  \right] } } {r \; p_{r}}  \label {v5ah}
\end{eqnarray}

For $N \geq 6$, the  process of determining the solution of
equation (\ref{app2}) goes tedious. Here, we argue that by
choosing $r$ sufficiently small, $v^{N-1}$ in equation
(\ref{app2}) can be neglected . For example, for $N=6$, $r=0.01$
and $\; p_{r}(r) = 1 + r^2 + r^3 + r^4 + r^5 + r^6 \;$, equation
(\ref{app2}) gives $\; v = 0.02714509001 \;$. With this, we
observe that for $N=6$ the term $\; v^{N-3} \; [ (N-1) (N-2)] / [2
r^2 p_{r} ] = 1.999999985 \;$ which is very close to $2$, while
the contribution from the term $\; v^{N-1}= 0.1473860672 \times
10^{-7} \;$ is negligible, hence, we can do way with the later
term. Therefore, for $N \geq 6$, we have
\begin{equation}
v_{Nah} = \left[ \frac{4 r^2 \; P_{r}} {(N-1) (N-2)} \right
]^{\frac{1} {N-3} }  \label {vnah}
\end{equation}

Now, we discuss the nature of central singularity  $r =0, t =
t_{s}(0)$ through the behaviour of causal paths emerging from the
singularity and reaching out to a far away observer in future or
otherwise. Using equation (\ref{v20}), the equation of apparent
horizon in (t, r) is written as follows
\begin{equation} t_{ah}(r) = t_{s}(r) - B(r) \label {ap04}
\end{equation}
where
\begin {equation}
B(r) = \int_{0}^{v_{ah}} \frac{v^{(N-3)/2} dv} { e^{\eta}
 \sqrt{ v^{N-3}  e^{2 \eta} b(r) +   v^{N-1} h(rv) + \frac{2 \; p_{r}\;
  ( 2 - v^{N-1})} {(N-1) (N-2)} } }   \label {ap05}
\end {equation}
Since the behaviour of apparent horizon is being studied close to
the central singularity at $r=0,R=0$, the upper limit of
integration in the above equation is small. Hence, the integrand
in above equation can be expanded in a power series in $v$ and
keeping only leading term, we obtain
\begin{equation}
t_{ah}(r) = t_{s_{o}} + r \chi_{p}(0) - \sqrt{ \frac{N-2} {N-1} }
\; (p_{r})^{\frac{-1} {2}} \; (v_{Nah})^ \frac{N-1} {2}  \label
{ap06}
\end{equation}
For $N=4,5$ and if $\chi_{p}(0)$ is non-zero positive, then in
above equation the second term dominates over the last negative
term so that the apparent horizon curve is increasing as we move
away from the singularity, therefore the singularity is naked. On
the other hand for $N \geq 6$, the negative term in equation
(\ref{ap06}) starts dominating and thus, advancing the time of
trapped surface formation. Hence for $N \geq 6$, we have a black
hole solution. These results coincide with the results obtained in
\cite{amrgpj} with radial pressure $p_{r}=0$.

These results can be verified by plotting curves
 for $\chi_{p}(0)=0.034, \; P_{r}=1 + r^2 + r^3
+r^4 + r^5 + r^6$ on small range of $r$ say $0$ to $0.075$ and
$tah(r)$ can be plotted with range of 0.9994 to 1.0024. Further,
with another value of $\chi_{p}(0)=0.5$, improving the range of
$r$ and $tah(r)$, we observe that curves for $N=4,5$ escape to
infinity.

\section* {V. Collapse with linear equation of state }
In this section, we aim at the question, whether the choice of
equation of state $ p = k \rho$, $k \in [0,1]$ contributes through
$k$ in the development of BH/NS phases. R. Goswami and P. S. Joshi
have studied the case of an isentropic perfect fluid with linear
equation of state in four dimensional spacetime, wherein, they
pointed that the occurrence of NS/BH evolving from regular initial
data depend on the choice of rest of the free function available
\cite{rgpj-2}. The stress-energy tensor for Type I field in a
diagonal form is given by \cite {he}
\begin{equation}
 T{^{t}_{t} } = -\rho, \; T{^{r}_{r} } =
 T{^{\theta}_{\theta}}=T{^{\phi}_{\phi} } = \; p \label {sem03}
\end{equation}
The quantities $\rho$ and $p=p_{r} =p_{\theta}$ are the energy
density and pressure respectively. We take the matter field to
satisfy weak energy condition which implies ${{\rho}{\geq}}0$;
${\rho+p{\geq}} 0$. Linear equation of state for perfect fluid is
\begin{equation}
p(t,r) = k \ \rho(t,r) \ where \  k \in [0, 1] \label {es04}
\end{equation}
 Einstein field equations for the metric
(1) are derived as
\begin{equation}
\rho = \frac{(N-2){\mathcal{F}}'}{2 R^{N-2}R'} \;  \;
 = - \frac{1} {k} \frac{(N-2)\dot{{\mathcal{F}}}}{2 R^{N-2} \dot{R}} \label {sfe05}
 \end{equation}
\begin{equation}
\nu' = - \frac{k}{k + 1} [ \ln(\rho)]' \label {sfe06}
\end{equation}
\begin{equation}
 R' \dot{G} - 2 \dot{R} G \ {\nu}' = 0  \label {sfe07}
\end{equation}
\begin{equation}
\hspace{.5in} G - H = 1 - \frac{{\mathcal{F}}}{R^{N-3}}. \label
{sfe08}
\end{equation}

As before, we have $R(t,r) = r \; v(t,r)$ and the choice of
physically reasonable initial data is maintained herein as in
section II. For further details, we refer the reader to \cite
{jd}. A general mass function for the cloud can be considered as
\begin{equation}
{\mathcal{F}}(t,r) = r^{N-1} \textit{M}(r,v) \label {sm12}
\end{equation}
where $\textit{M}(r,v)$ is regular and continuously twice
differentiable. Using equation (\ref {sm12}) in equations (\ref
{sfe05}), we obtain
\begin{equation}
\rho = \frac{N-2} {2} \times  \frac{(N-1) \textit{M} + r [
\textit{M}_{, r} + \textit{M}_{, v}
 \ v']}{ v^{N-2} ( v + r v')} \hspace{.1in}  =
  - \frac{(N-2) \textit{M}_{, v}} {2 k \ v^{N-2}}
 \label {sm13}
\end{equation}
Then as $ v \rightarrow 0 \; , \rho \rightarrow \infty $ and $p
\rightarrow \infty $ i.e. both the density and pressure blow up at
the singularity. At the initial epoch, the regular density
distribution takes the form
\begin{equation}
\rho_{o}(r) = \frac{N-2} {2} [r \textit{M}(r,1)_{,\ r} + (N-1)
\textit{M}(r,1) ] . \label {sie14}
\end{equation}
From equation (\ref {sm13}), it is clear that $\rho = \rho(r,v)$
and therefore, $v' = f(r,v)$. We rearrange equation (\ref{sm13})
as follows
\begin{equation}
(N-1) k \textit{M} + k r {\textit{M}}_{,\ r} + Z(r,v)
{\textit{M}}_{,\ v} = 0 \label {sv41}
\end{equation}
where
\begin{equation}
Z(r,v) = (k + 1) r v' + v  \label {sv15}
\end{equation}
Equation (\ref {sv41}) has general solution of the form,
\begin{equation}
{\mathcal{S}} (X,Y) = 0  \label {sv16}
\end{equation}
where $ X(r, v, \textit{M} )$ and $ Y(r, v, \textit{M}) $ are the
solutions of the system of equations  \cite{ve},
\begin{equation}
\frac{d \textit{M}} {(N-1) k } = \frac{dr} {k r } = \frac{ dv} {Z}
\label {sv17}
\end{equation}
Equation (\ref {sv16}) has many classes of solutions but only
those classes of solutions should be considered which satisfy
energy conditions, which are regular and gives $\rho \rightarrow
\infty$ as $v \rightarrow 0$. This means the energy conditions and
equation of state $ p = k \rho$ isolate the class of functions
$\textit{M}(r,v)$ so that the mass function ${\mathcal{F}}(t,r)$,
the metric function $ \nu (t,r)$ and the function $b(r)$ ( to
follow) evolve as the collapse begins according to the Type I
field equations.

On integrating equation (\ref {sfe06}), we obtain the general
 metric function,
\begin{equation}
\nu(r,v) = - \frac{k} {k + 1} [ \ln (\rho) ] \label {cv14}
\end{equation}
Define a suitably differentiable function $A(r, v)$,
 $A(r, v)_{,\ v} = {\nu'} / {R'}$ so that equation (\ref {cv14})
 takes the form
\begin{equation}
A(r,v)_{, \ v} = - \frac{k {\rho}'} {(k + 1) {\rho} R'} \label
{ecv15}
\end{equation}
At the initial epoch, we have
\begin{equation}
\left[A(r,v)_{,v} \right]_{v=1} = - \frac{k \ \rho_{o}'(r)} {(k +
1) \rho_{o}(r)}  \label {av16}
\end{equation}
whereas using equation (\ref {sm13}), the relation between the
function $\textit{M}$ and $A$, at all the epochs is given by
\begin{equation}
A(r,v)_{, v} R' = - \frac{k} {(k + 1)}  \left[ \ln \left( -
\frac{(N-2) {\textit{M}}_{, v}} {2 k \ v^{N-2}} \right) \right]'.
\label {tp17}
\end{equation}
From above, it is clear that $\textit{M}(r,v)$ is a decreasing
function with respect to $v$ and if we consider a smooth initial
profile for the density in such way that density gradient vanishes
at the center then we have $A(r,v) = r g_{o}(r,v)$ where
$g_{o}(r,v)$ is another suitably differentiable function.

  Also, the use of definition of $A(r,v)$ in equation (\ref {sfe07}) yields
\begin{equation}
G(t,r) = \textbf{d}(r) e^{2rA} \label {g18}
\end{equation}
where $ \textbf{d}(r) $ is another arbitrary continuously
differentiable function of $r$. We can write
\begin{equation}
\textbf{d}(r) = 1 + r^2 \textbf{b}(r).  \label {b17}
\end{equation}
where \textbf{b} is the energy distribution function for the
collapsing shells.
 Define a function $\textbf{h}(r,v)$ as
\begin{equation}
 \textbf{h}(r,v) = \frac{e^{2rA} - 1}{r^2}  \label {h19}
\end{equation}
and substituting this equation, together with (\ref {b17}) and
(\ref{g18}) in equation (\ref {sfe08}), we get
\begin{equation}
 v^{(N-3)/2} \; \dot{v} = - {\rho}^{-k/(k+1)} \sqrt{v^{N-3}
 \textbf{h}(r,v) + \textbf{b} v^{N-3} e^{2rA}
  + \textit{M}}. \label {sv20}
\end{equation}
where negative sign is chosen since, for the collapse, $ \dot{v} <
0 $. Integrating the above equation , we have
\begin{equation}
 t(v,r) = \int_{v}^{1}  \frac{ v^{(N-3)/2} dv}{{\rho}^{-k/(k+1)}
 \sqrt{v^{N-3} \textbf{h}(r,v) + \textbf{b} v^{N-3} e^{2rA}
  + \textit{M}}}
\hspace{.4in} \label {si21}
\end{equation}
In above equation, the variable $r$ is treated as a constant.
Expanding $t(v,r)$ around the center, we get
\begin{equation}
t(v,r) = t(v,0) + r \chi(v) + O(r^2) \label {t22}
\end{equation}
where the function
\begin{equation}
\chi(v) = \frac{dt}{dr} \Big{|}_{r=0} = - \frac{1}{2} \int_{v}^{1}
 \frac{v^{(N-3)/2} {\mathcal{B}}_{,r}(0,v) } {{\mathcal{B}}(0,v)^{3/2}} dv. \label {ch23}
\end{equation}
and
\begin{equation}
{\mathcal{B}}(r,v) = {\rho}^{-k/(k+1)} \sqrt{v^{N-3}
\textbf{h}(r,v) + \textbf{b} v^{N-3} e^{2rA} + \textit{M}(r,v)}
\label {chh24}
\end{equation}
\begin{equation}
t_{s_{o}} = \int_{0}^{1}  \frac{ v^{(N-3)/2} dv}
{{\mathcal{B}}(0,v)} \label {th25}
\end{equation}
\begin{equation}
t_{s}(r) \equiv t(0,r) = t_{s_{o}} + r \chi(0) + O(r^2) \label
{t26}
\end{equation}

Now, it is clearly seen that the value of $\chi(0)$ depends on the
free functions $\textbf{b}(0),{\mathcal{M}}(0,v)$ and $
\textbf{h}(0,v)$, which in turn, depend on the initial data at the
initial surface $ t=t_{i}$. Thus, a tangent to the singularity
curve $ t = t_{s_{o}} $ is completely determined by the given set
of density, pressure, velocity function $\nu$ and function
$\textbf{b}$. Further, from equation (\ref {sv20}), we can write
\begin{equation}
v^{(N-3)/2}  v' = \chi(v) {\mathcal{B}}(0,v) + O(r).  \label
{st24}
\end{equation}
Hence,
\begin{equation}
\chi(0) = \lim_{v \rightarrow 0} \chi(v) = \lim_{ v \rightarrow 0}
\frac{ v^{\frac{(N-3)} {2} } v'} {{\mathcal{B}}(0,v) }. \label
{scho25}
\end{equation}
The existence of the limit depends on the cumulative effect of all
the terms present in above equation but since $
{{\mathcal{B}}(0,v) }> 0$ as $v \rightarrow 0$ provided
$\textit{M}(0,0) > 0$ and $v \in [0,1]$, therefore positive or
negative sign of $\chi(0)$ absolutely depends on $v'$. So, if $v'
> 0$ then $\chi(0) > 0$ and otherwise. If $\chi(0) = 0$ then we
will have to take into account next higher order non-zero term in
the singularity curve equation and do a similar analysis. Thus, we
reckon that value of $\chi(0)$ depends on the initial data
profiles of density and pressure (through the free function
$\textit{M}(r,1)$), the energy function $\textbf{b}(r)$ and the
other free function $\textbf{h}(r,1)$ (for determination of the
metric function  $\psi$). Secondly, its sign depends on the nature
of gradient of $v$.

Now, we assume that these free functions to be at least $C^2$ and
satisfy other physical requirements. Let us choose the function
$M(r,v)$ as follows
\begin{eqnarray*}
\textit{M} (r,v) = 1 + (rv) + (rv)^2 + (rv)^3 + .......
\end{eqnarray*}
Now, the equation of apparent horizon $\; \frac{\mathcal{F}}
{R^{N-3}} =1 \;$ give rise to a polynomial equation in $v$
\begin{equation}
v^{N-3} - r v^{N-2} = r^2.   \label {spoly26}
\end{equation}
The apparent horizon $\; \textbf{V}_Nah \;$ for space dimensions
$N=4,5$ and $6$ are derived as follows
\begin{eqnarray*}
\textbf{V}_{4ah} = \frac{ 1 \pm \sqrt{ 1 - 4 r^3} } {2 r},
\end{eqnarray*}
next for $N=5$, we have
\begin{eqnarray*}
\textbf{A} = - 108 r^4 + 8 + 12 \sqrt{3}
 \; \sqrt {27 r^4 - 4 r^2}
 \end{eqnarray*}
 which is an imaginary term for sufficiently
  small values of $r$, $0 < r < \sqrt{4/27}$. Hence, apparent horizon
  may occur at some value of $r \geq \sqrt{4/27}$ for a chosen initial data
  of $\textit{M}(r,v)$.
  So, for such value of $r$, we can write
 \begin{eqnarray*}
 \textbf{V}_{5ah} = \frac {\textbf{A}^{1/3} } {6 r} + \frac{2/3} {r \textbf{A}^{1/3}}
 + \frac{1/3} {r}
\end{eqnarray*}
and
 \begin{eqnarray*}
\textbf{V}_{6ah} = \frac{1}{4r} + \frac{\sqrt{3} S} {12 r} +
{\frac{1} {12 r}} \sqrt{ - \frac{- 18 S T + 6 (12)^{1/3} r S
T^{2/3} + 24 (12)^{2/3} r^{4} S -18 \sqrt{3} T^{1/3} } {S T^{1/3}
}} \; \\
{where} \; T = r^2 ( 9 + \sqrt{- 768 r^{5} + 81}) \; \; {and} \;
\; S = \sqrt{  \frac{ 3 T^{1/3} + (212)^{1/3} r T^{2/3} + 8
(12)^{2/3}r^{4} } {T^{1/3}} }.
\end{eqnarray*}
The nature of central singularity  $r =0, t = t_{s}(0)$ as NS/BH
is studied through the behaviour of geodesics emanating from the
singularity to reach a asymptotic observer in future or otherwise.
Using equation (\ref{si21}), the equation of apparent horizon in
(t, r) is written as
\begin{equation} t_{ah}(r) = t_{s}(r) - {\textbf{B}_{1}}(r) \label {sap27}
\end{equation}
where
\begin {equation}
{\textbf{B}_{1}}(r) = \int_{0}^{V_{ah}} \frac{ v^{(N-3)/2} \;
dv}{{\rho}^{-k/(k+1)}
 \sqrt{v^{N-3} \textbf{h}(r,v) + \textbf{b} v^{N-3} e^{2rA}
  + \textit{M}}}   \label {sap28}
\end {equation}
The upper limit of integration in the above equation is small as
discussed earlier. Using the value of $\rho$ from equation
(\ref{sm13}) and expanding the terms in the integrand in a power
series in $v$ and keeping only leading term, we obtain
\begin{equation}
t_{ah}(r) = t_{s}(r) - \frac{[ (N-1) (N-2)]^{m}} {2^{m}}
\int_{0}^{V_{ah}}  {\frac{  v^{ (2m-N-3)/2 } \;  dv} { (1 - rv)^{
-(1-2m)/2 } }  } \label {sap29}
\end{equation}
where we have put $m =k/(k+1)$ and $ 0 \leq m \leq {1/2} $.
Further, since $r$ is sufficiently small, therefore, we can do
away with the terms $O(rv)^{2}$, hence for $N \geq 4$, we obtain
\begin{equation}
t_{ah}(r) = t_{s_{0}} + r \chi{0} + \frac{[ (N-1) (N-2)]^{m}}
{2^{m}} \left[ \frac{2} {N+1-2m} - \frac{  r (1-2m) V_{Nah} }
{N-2m -1} \right] (V_{Nah})^{-  (N+1-2m)/2}
  \label {sap30}
\end{equation}
The second term in above equation dominates over the last for
sufficiently small values of $r$ and for any value of $k$ (the
parameter of equation of state) and independent of any choice of
dimension of the spacetime. Hence, causal paths emerging from the
singularity reach out to an asymptotic  observer at infinity, thus
naked singularities appear in all dimensions for sufficiently
small values of $r$. The results for dust case ie. with $m=0$
agree with the results obtained in \cite{kd}.

\section* {VI. Conclusion}
We have considered the radial pressure $p_{r}$ as a function of
$r$, due to which the mass function $F(t,r)$ becomes the function
of both $r$ and $v$, while the tangential pressure remains
arbitrary through out the analysis. Through the study of apparent
horizon, it is found that NS forms at the center of the collapsing
cloud for dimensions $N=4$ and $N=5$, whereas for $N \geq 6$ there
exists a critical value of $\chi(0)$ below which a BH solution
persists. This statement supports the understanding that gravity
may be a higher dimensional interaction. This generalizes the
results obtained in \cite{amrgpj}.

Next, we have analyzed the effects of the equation of state on the
formation of apparent horizon in continual gravitational collapse,
the parameter $k$ involves in determination of the curve
$t=t_{ah}(r)$ and assist in strengthening the out going causal
path to reach out to an asymptotic observer at infinity. It is
analyzed that naked singularities exist in all the dimensions.
$k=0$ is the usual case of dust collapse.

\end{document}